\documentclass[aps,prd,floatfix,showpacs,superscriptaddress,nofootinbib,10pt]{revtex4-2}
\usepackage{graphicx}
\usepackage{amsmath,bbold}
\usepackage{bm,cancel}
\usepackage{slashed}
\usepackage{xcolor,braket}
\newcommand{\be}{\begin{equation}}
\newcommand{\ee}{\end{equation}}

\usepackage[colorinlistoftodos,prependcaption,textsize=small]{todonotes}
\usepackage[normalem]{ulem}

\makeatletter
\newcommand*{\encircled}[1]{\relax\ifmmode\mathpalette\@encircled@math{#1}\else\@encircled{#1}\fi}
\newcommand*{\@encircled@math}[2]{\@encircled{$\m@th#1#2$}}
\newcommand*{\@encircled}[1]{%
  \tikz[baseline,anchor=base]{\node[draw,circle,outer sep=0pt,inner sep=.2ex] {#1};}}
\makeatother

\newcommand{\CCE}{Centro de Ciencias Exactas, Universidad del Bio-Bio, Avda.~Andr\'es Bello 720, Casilla 447, Chill\'an, Chile.}
\newcommand{\UBB}{Departamento de Ciencias B\'asicas, Facultad de Ciencias, Universidad del Bio-Bio, Avda.~Andr\'es Bello 720,
Casilla 447, Chill\'an, Chile}
\newcommand{\kulak}{KU Leuven Campus Kortrijk -- Kulak, Department of Physics, Etienne Sabbelaan 53 bus 7657, 8500 Kortrijk, Belgium}
\newcommand{\ughent}{Ghent University, Department of Physics and Astronomy, Krijgslaan 281-S9, 9000 Gent, Belgium}
\newcommand{\unicamp}{Universidade Estadual de Campinas - Instituto de F\'isica ``Gleb Wataghin''
Rua S\'ergio Buarque de Holanda, 777, CEP 13083-859 - Campinas, SP, Brazil}
\newcommand{\unesp}{Instituto de F\'{i}sica Te\'{o}rica, Universidade Estadual Paulista, Rua Dr.~Bento Teobaldo Ferraz, 271 - Bloco II, 01140-070 S\~{a}o Paulo, SP, Brazil}

\begin{document}

\title{Half-integer anomalous currents in 2D materials from a QFT viewpoint}

\author{David Dudal}\affiliation{\kulak}\affiliation{\ughent}
\author{Filipe Matusalem}\affiliation{\unicamp}
\author{Ana J\'{u}lia Mizher}\affiliation{\CCE}
\affiliation{\unesp}
\author{Alexandre Reily Rocha}\affiliation{\unesp}
\author{Cristian Villavicencio}\affiliation{\CCE}\affiliation{\UBB}


\begin{abstract}
Charge carriers in Dirac/Weyl semi-metals exhibit a relativistic-like behavior. In this work we propose a novel type of intrinsic half-integer Quantum Hall effect in 2D materials, thereby also offering a topological protection mechanism for the current. Its existence is rooted in the 2D parity anomaly, without any need for a perpendicular magnetic field. We conjecture that it may occur in disturbed honeycomb lattices where both spin degeneracy and time reversal symmetry are broken. These configurations harbor two distinct gap-opening mechanisms that, when occurring simultaneously, drive slightly different gaps in each valley, causing a net anomalous conductivity when the chemical potential is tuned to be between the distinct gaps. Some examples of promising material setups that fulfill the prerequisites of our proposal are also listed to motivate looking for the effect at the numerical and experimental level.
\end{abstract}

\maketitle

\section{Introduction}
Recently discovered Weyl and Dirac semi-metals allow to make concrete connections between relativistic quantum field theory phenomena and condensed matter physics due to the linear dispersion in these materials, a feature that is typically associated to relativistic particles \cite{relativistic}. Exactly due to this relativistic-like behavior, combined with the low energies associated, it is possible to construct table top setups to explore certain phenomena proposed to happen in particle physics but that are too hard to observe due to the complexity of the experimental apparatus and data analysis in this field. Remarkable examples are the Klein paradox \cite{klein} and the Zitterbewegung \cite{Zitterbewegung}.

At the same time, two-dimensional materials such as graphene \cite{Novoselov2007,castroneto} and transition metal dichalcogenides form a new family of materials, interesting from the fundamental physics as well as applications viewpoint \cite{rev1,rev2}. A low energy tight binding description of planar Dirac materials arranged in a honeycomb lattice yields a Clifford algebra associated to the sublattice, or pseudo-spin degree of freedom \cite{Gusynin:2007ix,katsnelson}. Despite that in two dimensions it is not possible to define the chiral operator $\gamma^5$, it is possible to combine the spinors associated to each Dirac point in a four-component spinor containing sublattice and valley degrees of freedom. This corresponds to a merging of the two representations in a new, reducible, four-component one \cite{deJesusAnguianoGalicia:2005ta}. In this fashion, the usual $4\times 4$ Dirac matrices emerge \cite{Gusynin:2007ix} and it is possible to define an operator analogous to the $(3+1)$-dimensional chirality operator $\gamma^5$ which commutes with the Hamiltonian, indicating a conserved quantum number. The eigenstates from the Hamiltonian, $\psi_k$ and $\psi_{k'}$, are eigenstates of this operator with eigenvalues $\pm 1$. This valley degree of freedom is sometimes also coined pseudo-chirality.

Although the pseudo-spin and pseudo-chirality share exactly the same group theory with the spin and chirality respectively, this correspondence has mostly  been considered as a mere analog. Nevertheless it was shown that the situation may drastically change when the sublattice symmetry is broken and consequently a gap is opened \cite{Mecklenburg,photonic}. In this case, experimental observations indicate there is an angular momentum associated with the pseudo-spin, giving it a remarkable physical meaning.

\begin{figure}
\includegraphics[width=0.5\textwidth]{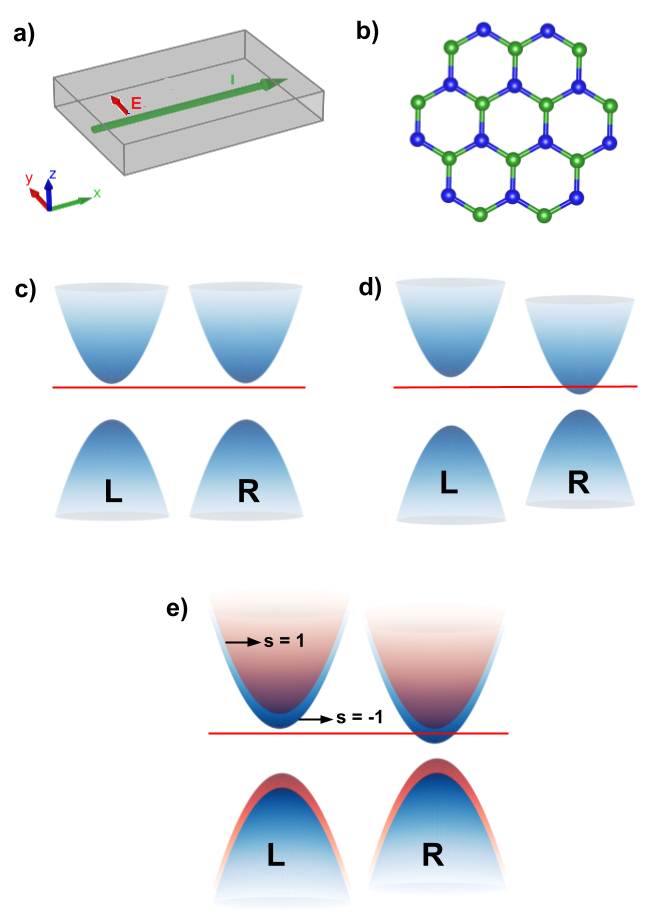}
\caption{(a) The A-QHE generates an electric current perpendicular to the external electric field; (b) Honeycomb lattice with broken sublattice symmetry; (c) Dirac cones with simple gap, analogous to the chirally symmetric case; (d) Valley asymmetry corresponding to imbalance in chirality; (e) Schematic representation of the band structure when a spin split is present besides valley asymmetry.}
\label{ABCDE}
\end{figure}

Our current goal is to provide theoretical evidence, from an effective quantum field theory viewpoint, for a topologically protected half-integer Anomalous Quantum Hall Effect (A-QHE) in suitably gapped 2D materials. Concretely, we will identify a current of the type shown in Fig.~\ref{ABCDE}(a), with associated quantized conductivity equalling $\frac{e^2}{2\hbar}$. The A-QHE has triggered an intense research activity \cite{Haldane:1988zza,aqhe,V2O3}, since it would be intrinsic to the material without the need for a (strong) external magnetic field as for the normal QHE, and when a 2D material with A-QHE is put in heterostructure with a superconducting material, the ensuing proximity effect could
lead to the formation of the elusive Majorana fermions \cite{Qi1,Qi2}, with great potential for (topologically protected
against decoherence) qubits and thence quantum computation therewith. Interestingly enough, the inspiration behind also comes from a quantum field theory (QFT) viewpoint, see \cite{Jackiw:1981ee}.  The intrinsic nature of the A-QHE is important in the latter context, as strong magnetic fields required for the ``normal'' QHE (which is non-intrinsic), would be destructive for the superconductor.

\section{Theoretical setup from a QFT viewpoint}
\subsection{Setting the stage}
We explore a honeycomb lattice with broken sublattice symmetry, as represented in Fig.~\ref{ABCDE}(b). It has been shown that the difference in the energy of the electrons belonging to different sublattices can be parameterized at the Hamiltonian level via an effective standard fermion mass for two two-component spinors, $\pm m$, with opposite sign for each valley \cite{Semenoff,Redlich:1983kn}. This induces a gap opening that preserves time reversal symmetry. In this case both valleys present symmetric gaps as represented in Fig.~\ref{ABCDE}(c). It is also known that the anomalous {\sf T}-odd piece of the electric current generated by an external field acting on this system is related to a topological Chern-Simons term \cite{Niemi:1983rq}. In other words, it depends solely on the sign of the mass term and not on its magnitude. In presence of an external gauge field in two spatial dimensions, one gets
\be
j^i(x)=\frac{e^2}{4\pi}\frac{m}{|m|}\varepsilon^{ijk}F_{jk} (x)=\frac{e^2}{4\pi}\text{sign}(m)\varepsilon^{ijk}F_{jk} (x),
\ee
indicating that the net current vanishes when one sums up the contribution of both valleys \cite{Semenoff}.

In this work we add two new ingredients to the case previously considered. First of all, we allow for arbitrary values of the gaps associated to each Dirac point, $m_k$ and $m_{k'}$, enabling configurations where $m_k\neq m_{k'}$. This corresponds to a band structure where the two valleys have asymmetric gaps, as represented in Fig.~\ref{ABCDE}(d). Secondly, we include a chemical potential in order to explore certain regions of the band structure. The reason for this flexibility and how we calibrate the chemical potential will soon become clear.

For pure honeycomb lattices composed by one type of atom per lattice site, such as pristine graphene, the Lagrangian obtained from the continuum limit of the tight-binding model is very similar (apart from the Fermi velocity breaking Lorentz invariance) to the fermion sector of Quantum Electrodynamics in two space and one time dimensions \cite{Gusynin:2007ix,katsnelson},
\begin{eqnarray}
\mathcal{L}&=&\sum_s \bar{\psi}_s\left(i\gamma^0\hbar D_t +i\hbar v_F\gamma^xD_x + i\hbar v_F \gamma^y D_y \right)\psi_s.
\label{eq:lagrangian}
\end{eqnarray}
Here the index $s$ labels the (electron) spin and the covariant derivative $D_\alpha=\partial_\alpha-(ie/\hbar c) A_\alpha$. $A_\alpha$ is the gauge field associated to the electromagnetic interaction and the fields $\psi_s\equiv \psi_s(t,{\bf r}) $ denote four-component spinors that account for both valley and pseudospin index.

We choose to work in the Weyl basis,
\be
\gamma^0=\left(
           \begin{array}{cc}
             0 & \mathbb{1}_2 \\
             \mathbb{1}_2 &0 \\
           \end{array}
         \right),\qquad \gamma^{x,y,z}=\left(
                                   \begin{array}{cc}
                                     0 & \sigma^{x,y,z} \\
                                     -\sigma^{x,y,z} & 0 \\
                                   \end{array}
                                 \right),\qquad \gamma^5=\left(
           \begin{array}{cc}
             -\mathbb{1}_2 &0  \\
             0&\mathbb{1}_2  \\
           \end{array}
         \right).
\ee

To make the valley degree of freedom more transparent we rewrite our Lagrangian using the pseudo-chiral projections. Decomposing the fermion field as
\begin{equation}
\psi=\psi_k + \psi_{k'}=\frac{1}{2}\left(1+ \gamma^5 \right)\psi + \frac{1}{2}\left(1- \gamma^5 \right)\psi,
\end{equation}
we can split the Lagrangian in Eq.~(\ref{eq:lagrangian}) in two terms, associated to left and right pseudo-chiralities. We introduce untied gaps for each pseudo-chirality, allowing each projection to have different masses. The resulting Lagrangian is \cite{Mizher:2013kza,Mizher:2018dtf}
\begin{eqnarray}
\mathcal{L}&=& \sum_{s,\chi=k,k'}\bar{\psi}_{\chi,s}[i\gamma^0\hbar\partial_t +\mu\gamma^0+i\hbar v_F\gamma^xD_x + i\hbar v_F \gamma^y D_y+m_{s,\chi}\gamma^z]\psi_{\chi,s},
\label{eq:chirallagrangian}
\end{eqnarray}
where we included the chemical potential, $\mu$, and a more general mass gap structure with $m_{s,k}^2\neq m_{s,k'}^2$. Without loss of generality, we may assume $m_km_k'>0$.

\subsection{The fermion mass sector in two-component spinor language}
For later usage, it is useful to rewrite the Lagrangian \eqref{eq:chirallagrangian} a bit. The mass terms are written in four-component spinor language in the pseudo-chiral basis, meaning that (we suppress the spin index $s$ for now.)
\be
\psi_k=\left(
  \begin{array}{c}
    0 \\
    \psi_+ \\
  \end{array}
\right),\qquad \psi_{k'}=\left(
  \begin{array}{c}
    \psi_- \\
    0 \\
  \end{array}
\right)
\ee
using two two-component spinors $\psi_\pm$.  The corresponding mass sector
\be
S_{\mathrm{mass}}=\int d^3x \left(\bar\psi_k (\mu\gamma^0+m_k\gamma^z)\psi_k+ \bar\psi_{k'} (\mu\gamma^0+m_{k'}\gamma^z)\psi_{k'} \right)
\ee
can be rewritten as
\be
S_{\mathrm{mass}}=\int d^3x \left(\bar\psi_+ (\mu\sigma^z+m_k)\psi_+ + \bar\psi_- (\mu\sigma^z-m_{k'})\psi_- \right),
\ee
keeping in mind that in $(2+1)$-dimensions, $\sigma^z$ plays the role of $\gamma^0$ in an appropriate basis \cite{Dunne:1998qy}. This shows that we actually have two two-components spinors with opposite sign standard masses. This plays a pivotal role in our analysis.

\subsection{Linear response theory applied to Reduced Quantum Electrodynamics}

Linear response theory investigates the reaction of a system under a small external influence assuming that this reaction can be studied at linear order in the external stimulus. Here we apply this procedure to calculate the conductivity of the just introduced continuum quantum field theory description of a material sample organized in a $(2+1)$-dimensional honeycomb lattice when a constant external electric field is applied in-plane, let's say $\vec E= E \vec e_y$. The electric current we are interested in is then given by
\be
\langle \vec j \rangle=\sigma_{xy} E \vec e_x,
\ee
also in-plane but perpendicular to the applied $\vec E$.

A certain care is needed since the underlying system is actually a mixed-dimensional theory: the charge carriers are constrained to a plane while the gauge fields live in a bulk. The theory to describe such systems has been developed previously and is known as pseudo-Quantum Electrodynamics (PQED) \cite{PQED,PQED2} or reduced-Quantum Electrodynamics (RQED) \cite{RQED}. Here we present in detail how to introduce the interaction with the external electric field. For the sake of clarity, we refer in this subsection to the dimension of the system using the usual quantum field theory notation ($n$-space~$~+1$-time)-dimension, in contradistinction with the condensed matter conventional notation used elsewhere in the manuscript.

The Kubo formalism has been successfully applied to several transport phenomena in Dirac/Weyl materials \cite{kubo_Hall,2-loop}. To extract the DC conductivity, we will benefit of some of the special properties of our mixed-dimensional system. It is a common trick to temporarily consider a time-varying electric field, and ultimately the DC limit $\omega\to0$ will yield the desired constant field.  The external (classical) electric field, and conveniently chosen vector potential,  read
\be\label{b}
\vec{E}=E e^{ - i\omega t} \vec{e}_y\,,\qquad \vec{a}=-\frac{iE}{\omega}e^{-i\omega t } \vec{e}_y
\ee
in natural units where $\hbar=c=1$. Notice that this field configuration does not solve the classical Maxwell equations and cannot be considered a physical disturbance, unless for $\omega\to0$. We can introduce, however, a magnetic field in order to solve the Maxwell equations, but it will vanish in the limit $\omega\to 0$, so we can omit it.

Having introduced the classical background, we proceed with the action in $(3+1)$-dimensions, given by
\be
S=\frac{1}{4}\int d^4 x F_{\mu\nu}^2 + \int d^3 x  \bar{\psi}(\vec{x},0)i\slashed{D}\psi(\vec{x},0)+\rm{fermion~mass\  terms}+\rm{gauge\ fixing},
\label{action}
\ee
where the vector $\vec{x}=(t,x,y)$ and the covariant derivative $\slashed{D}=(\partial_t -ie A_0)\gamma^0  +  v_F(\partial_x-ieA_x)\gamma^x+ v_F(\partial_y-ieA_y)\gamma^y$, since the (fermion) electronic degrees of freedom propagate on the $z=0$ surface.

One can check the full gauge invariance of the $(3+1)$-dimensional action in Eq.~(\ref{action}), applying
\begin{eqnarray} \nonumber
\psi(\vec{x},0)\rightarrow e^{-ie\alpha(x)}\psi(\vec{x},0),\quad
\bar{\psi}(\vec{x},0)\rightarrow e^{ie\alpha(x)}\bar{\psi}(\vec{x},0),\quad
A_\mu(x) \rightarrow A_\mu(x) -\partial_\mu \alpha(x),
\end{eqnarray}
with 4-vector $x= (\vec{x},z)$. We tacitly assume so-called absolute boundary conditions \cite{Vassilevich:2003xt}, $\left.A_z\right|_{z=0}=0$, $\left.\partial_z \alpha\right|_{z=0}=0$, $ \left.\partial_z A_{0,x,y}^{z>0}\right|_{z=0}-\left.\partial_z A_{0,x,y}^{z<0}\right|_{z=0}=j_{0,x,y}$, as also used in \cite{Dudal:2018pta,Herzog:2017xha}. Notice that these are consistent with using the Landau gauge.

The Noether current is easily identified as
\be
 j_\mu(\vec{x}) = e\bar{\psi}(\vec{x},0)\gamma_\mu \psi(\vec{x},0).
\ee
This current is gauge invariant and conserved on-shell, as expected. It is also independent from the $z$-coordinate. The third component $j_z$ can be written down formally, but it does not correspond to the transport of any physical electric charge, as can be easily checked using the integrated Gauss'~law. So effectively, $j_z=0$.

We notice that $j_\mu$ has mass dimension~$2$, as corresponding to a $(2+1)$-dimensional current. Let us introduce its $(3+1)$-dimensional version,
\be
J_\mu(x) = j_\mu(\vec{x})\delta(z)
\ee
to properly disturb the action $S$ with
\begin{eqnarray} \nonumber
\int d^4 x J_\mu (x) a_\mu(x)= \int d^4 x  j_y(\vec{x})\frac{-iE}{\omega} e^{-i\omega t}\delta(z)
= \int d^3 x  j_y(\vec{x})\frac{-iE}{\omega} e^{-i\omega t}
\end{eqnarray}
to set up the appropriate Kubo linear response theory.

The (still mixed-dimensional) perturbed action can thus be written as
\begin{eqnarray}
\widetilde S&=&\frac{1}{4}\int d^4 x F_{\mu\nu}^2 + \int d^3 x \bar{\psi}(\vec{x},0)(i\slashed D) \psi(\vec{x},0) + \int d^4 x J_\mu(x) a_\mu(x)+\rm{fermion\ masses}+\rm{gauge\ fixing}.
\label{action2}
\end{eqnarray}
Following e.g.~\cite{Dudal:2018pta} and transforming to momentum (Fourier) space, it can be shown that the action in Eq.~(\ref{action2}) can be reduced to an equivalent, be it purely $(2+1)$-dimensional, action
\begin{eqnarray}
\widetilde S&=&\frac{1}{2}\int d^3 x F_{\mu\nu} \frac{1}{\sqrt{-\partial^2}}F_{\mu\nu} + \int d^3 x \bar{\psi} i \slashed{D}\psi + \int d^3 x j_y (\vec{x}) \left(\frac{-iE}{\omega} e^{-i \omega t}\right)+\rm{fermion\ masses}+\rm{\ gauge\ fixing}.
\end{eqnarray}
We recognize here RQED, supplemented with an external disturbance. Doing so, we can now depart from this consistent $(2+1)$-dimensional action to extract the DC conductivity via a Kubo relation. There is no more reference to the obsolete $z$-direction. This also makes clear why we chose the vector potential \eqref{b}, which is by no means unique, as this choice allows for a full dimensional reduction of the relevant dynamics.

Closely following \cite{Tong:2016kpv}, we have a disturbed Hamiltonian given by $\Delta H=\vec j \cdot \vec{a}$, from now on always assuming $z=0$. The expectation value of the current to leading order in $\vec{a}$ becomes
\begin{eqnarray}
\braket{\vec j(t)}= \frac{i}{\hbar}  \bra{0} \int_{-\infty}^{t} d\tau [\Delta H(\tau),\vec j(t)]\ket{0}.
\end{eqnarray}
or concretely, after using time translational invariance, we get
 the Kubo relation for the DC anomalous conductivity,
\be\label{sigma}
 \sigma_{xy} = \lim_{\omega\to0}\frac{1}{\hbar\omega}\left\{ \int_{0}^{\infty} d\tau e^{i\omega\tau}  \bra{0} [j_y(0),j_x(\tau)]\ket{0}\right\}.
\ee
As we are interested in the $\omega\to0$ limit, it will be sufficient to evaluate the integral in Eq.~\eqref{sigma} up to $\mathcal{O}(\omega)$. As the integrand in the r.h.s.~of Eq.~\eqref{sigma} is actually the retarded correlator, we can use the fact that retarded and Euclidean self-energy correlator coincide at $\omega=0$. We may use \cite[Eq.~(2.11)]{Meyer:2010ii} to find they do so up to $\mathcal{O}(\omega^2)$, an identification valid since the Euclidean photon self-energy $\Pi^{ij}$ has no pole at zero frequency (see Eq.~\eqref{piij}). In fact, from Eq.~\eqref{pi3} it follows the spectral function will only be non-vanishing from a non-zero threshold onwards, exactly due to the infrared being protected by the mass scale.

\subsection{The RQED self-energy at zero momentum}
Assuming first a standard Dirac mass $m$ for a single two-component spinor and a chemical potential $\mu$, it can be shown that the low-energy limit of the induced $\sf T$-odd part in the RQED photon self-energy is one-loop exact \cite{Dudal:2018mms}. This is the Coleman-Hill theorem \cite{Coleman:1985zi} generalized to RQED, resulting in a ``topological'' Chern-Simons mass term in the effective action. Moreover, that relevant contribution is also independent of the Fermi velocity $v_F$ \cite{Dudal:2018mms}. At one-loop the (transverse and gauge invariant) polarization tensor is only depending on fermion propagators, which are just the same as in usual planar QED.

The  required (Euclidean) one-loop photon self-energy---which corresponds to the current-current correlator---reads
\begin{equation}\label{pi1}
  \tilde \Pi^{ij}(p)= e^2v_F^2 \int \frac{d^3q}{(2\pi)^3} \textrm{Tr}\left(\gamma^i \frac{\widehat{\slashed{p}+\slashed{q}}-m}{(\widehat{p+q})^2+m^2}\gamma^j \frac{\widehat \slashed q-m}{\widehat q^2+m^2}\right)
\end{equation}
where in general $\widehat k = (k_0+i\mu, v_Fk_x, v_Fk_y)$.

A non-vanishing contribution to $ \tilde\Pi^{xy}$ can only arise from the $\sf T$-odd piece of the above integral, a piece which is necessarily proportional to the Levi-Civita tensor $\varepsilon^{ijk}$. The latter comes exclusively from the product of three $\gamma$-matrices, based on $\textrm{Tr}(\gamma^i \gamma^j \gamma^k)=-2\varepsilon^{ijk}$. So we get, with $p_0\equiv \omega$,
\begin{equation}\label{pi2}
   \tilde\Pi^{ij}(\vec p)=\varepsilon^{ijk} p_k \pi(p)
\end{equation}
where
\begin{equation}\label{pi3}
   \pi(\vec p)=2me^2v_F^2\int\frac{d^3q}{(2\pi)^3}\frac{1}{(\widehat{q +  p})^2+m^2}\frac{1}{\widehat q^2+m^2}.
\end{equation}
Given the explicit $p_k$ in front of expression \eqref{pi2} and the eventual $\omega\to0$ limit, it is sufficient for our purposes to compute $\pi(0)$,
\begin{equation}\label{pi4}
  \pi(0)=2me^2v_F^2\int\frac{d^3q}{(2\pi)^3}\frac{1}{((q_0+i\mu)^2+v_F^2q_x^2+v_F^2q_y^2+m^2)^2}=
  e^2\frac{m}{4\pi^2}\int dq_0\frac{1}{(q_0+i\mu)^2+m^2}=\frac{e^2}{4\pi}\frac{m}{|m|}\theta(m^2-\mu^2)\,,
\end{equation}
the last step based on the residue theorem, the integral vanishes if $m^2<\mu^2$ as the contour can then be chosen to not encircle any of the poles occurring at $p_0=-i\mu\pm i|m|$.  Putting everything together, we actually have
\be\label{piij}
\tilde\Pi^{yx}(\vec p) = -\frac{m}{|m|}\frac{e^2}{4\pi} \theta(m^2-\mu^2)\varepsilon^{yx0}\omega + \mathcal{O}(p^2).
\ee
This result coincides with that of e.g.~\cite{Niemi:1984vz,Poppitz:1990nv,Sisakian:1996cb,Zeitlin:1996ys}, where the temperature $T\to0$ limit is understood whenever necessary. Let us  refer the interested reader to e.g.~\cite{nieuw1,nieuw2} for alternative derivations for several types of Quantum Hall conductivities.

To avoid any doubts about the DC limit, we find it useful to explicit verify that our result is independent on how the zero momentum limit is taken, namely $p_0=0$, $\lim p_{x,y}\to0$ vs.~$p_{x,y}=0$, $\lim p_{0}\to0$. Adopting the Feynman trick,
\begin{equation}\label{pi3bis}
   \pi(\vec p)=2me^2v_F^2\int\frac{d^3q}{(2\pi)^3}\int_0^1d\alpha\frac{1}{\left[\alpha((\widehat{q +  p})^2+m^2)+(1-\alpha)(\widehat q^2+m^2)\right]^2}
\end{equation}
or, after substituting $\ell_0=q_0+\alpha p_0$, $\ell_{x,y}=v_F(q_{x,y}+\alpha p_{x,y})$,
\begin{equation}\label{pi3tris}
   \pi(\vec p)=2me^2\int_0^1d\alpha\int\frac{d^3\ell}{(2\pi)^3}\frac{1}{((\ell_0+i\mu)^2+\ell_x^2+\ell_y^2+m^2+\alpha(1-\alpha)\vec p^2)^2}.
\end{equation}
As the net dependence is clearly on $\vec p^2=p_0^2+v_F^2p_x^2+v_F^2p_y^2$, from the above expression it is already obvious the aforementioned limits do commute. For the momentum integration, we can follow exactly the same procedure as in the $\vec p^2=0$ case. So, after integrating over the Feynman parameter, we end up with
\begin{equation}\label{pi3quattro}
   \pi(\vec p)=\left\{
                 \begin{array}{ll}
                   \frac{e^2}{2\pi}\frac{|m|}{p}\textrm{ArcCot}\left(2\frac{|m|}{|p|}\right) & \hbox{if}~m^2\geq\mu^2 \\
                   \frac{e^2}{2\pi}\frac{|m|}{p}\textrm{ArcCot}\left(2\frac{|\mu|}{\sqrt{4m^2-4\mu^2+\vec p^2}}\right) & \hbox{if}~m^2<\mu^2~\&~\vec p^2>4\mu^2-4m^2 \\
                   0 & \hbox{otherwise.}
                 \end{array}
               \right.
\end{equation}
It can be verified that in the $\mu\to0$ limit, this result coincides with the one quoted in e.g.~\cite{Dunne:1998qy}. Evidently, Eq.~\eqref{pi3quattro} reduces to Eq.~\eqref{pi4} for $\vec p\to0$.

\subsection{The DC conductivity itself}
Putting everything back together, we get from Eqs.~\eqref{sigma}, \eqref{piij} that
\be
\sigma = \lim_{\omega\to0}\frac{1}{\hbar \omega} \Pi^{yx} =-\lim_{\omega\to0} \frac{1}{\hbar\omega} \frac{m}{|m|}\frac{e^2}{4\pi} \theta(m^2-\mu^2)\varepsilon_{yx0}\omega= \frac{e^2}{4\pi\hbar}\frac{m}{|m|}\theta(m^2-\mu^2)
\ee
for a single massive two-component spinor. This DC conductivity will receive no further corrections, given the one-loop exactness of $\Pi^{yx}(\omega)$ for vanishing frequency. This is suggestive of the fact the conductivity might have a topological origin, something that will be discussed further in the next Subsection.

To derive our final expression for the anomalous conductivity, in the case that only a single spin-band contributes, albeit two different two-component spinors (one per Dirac point),
\be\label{naam}
\sigma_{xy} = \frac{e^2}{4\pi\hbar}\left(\frac{m_k}{|m_k|}\theta(m_k^2-\mu^2)-\frac{m_{k'}}{|m_{k'}|}\theta(m_{k'}^2-\mu^2)\right).
\ee
Assuming now that $m_k>0$, $m_{k'}>0$ and $m_{k'}^2<\mu^2< m_{k}^2$, the predicted value for the conductivity becomes
\be\label{final}
\sigma_{xy} = \frac{e^2}{2h},
\ee
which corresponds to a \emph{half-integer} anomalous quantum Hall conductivity capable of sustaining a current along an externally applied electric field. A pivotal role is played here by assuming the chemical potential to be in the gap between the two gaps, otherwise we would get a net cancellation of the two terms in Eq.~\eqref{naam} yet again.

Reintroducing the spin degree of freedom, we have
\begin{eqnarray}
\sigma_{xy}&=&\sum_s \frac{e^2}{4\pi}\Big[\frac{m_{s,k}}{|m_{s,k}|}\theta(m_{s,k}^2-\mu^2)-\frac{m_{s,k'}}{|m_{s,k'}|}\theta(m_{s,k'}^2-\mu^2)\Big],
\label{eq:conductivity}
\end{eqnarray}
where $m_{s,k}$ and $m_{s,k'}$ are now the gaps associated to each valley, per spin choice.

This is our main result. We note that for the configuration considered in previous works \cite{Mecklenburg,Semenoff}, where $|m_{s,k}|=|m_{s,k'}|$, the net current vanishes as expected, but let us consider the case where the gaps differ, represented in Fig.~\ref{ABCDE}(d). Then the system does enjoy neither a space inversion nor time reversal invariance, implying an uplifting of the pseudo-chiral fermions degeneracy. The most immediate way to achieve such a configuration is to work simultaneously with a double gap opening mechanism. On top of the inversion symmetry breaking mass term, $\delta \tau_z \otimes \sigma_z$---with $\tau$ designating the valley whilst $\sigma$ relating to sublattice---discussed previously \cite{Semenoff},
we propose to add an interaction capable of inducing a time reversal symmetry breaking. This corresponds to a mass term where the sign of the masses remain the same in both valleys, $\Delta \tau_0 \otimes \sigma_z$. A configuration subject to both symmetry breaking mechanisms will lead to different gaps in each valley as required.

The coupling of the masses to spin must be chosen carefully, otherwise currents associated to different spins cancel each other. First of all, we require that a spin flip causes a global change of sign, not a relative one. This is crucial otherwise for each spin a different valley will be favored, inflicting a cancellation. That is the reason a Kane-Mele type of configuration \cite{kane_mele} does not satisfy our conditions. In order to avoid this situation we choose to add the following mass couplings
\begin{eqnarray} \nonumber
m_{s,k}\sigma_z&=&s(\Delta  +\delta)\sigma_z,\qquad
m_{s,k'}\sigma_z=s(\Delta - \delta)\sigma_z,
\label{eq:spin_coupling}
\end{eqnarray}
to the Lagrangian, with $\Delta > \delta$.

Notice however that a flip in spin induces a flip in mass sign. We can check from Eq.~(\ref{eq:conductivity}) that this also would induce a cancellation between currents belonging to the same valley with different spin. We will consider this observation below, when we seek for suitable materials that fulfill our conditions.

Synthesizing the discussion above, a planar material capable to support our proposed half-integer A-QHE must ({\it i}) contain Dirac points ({\it ii}) present inversion symmetry breaking ({\it iii}) present time reversal symmetry breaking and ({\it iv}) present a split in spin bands.

If these conditions are met, we thus predict from Eq.~\eqref{eq:conductivity} a DC current given by
\begin{equation}\label{finaalresultaat}
\vec j = \sigma_{xy} E \vec e_y,\qquad \sigma_{xy} = \frac{1}{2}\frac{e^2}{h}.
\end{equation}
where we reintroduced the units.

Compared to the Haldane model \cite{Haldane:1988zza}, there are two differences. We consider two symmetry breakings---apart from a spin polarization to lift the spin degeneracy, which is also present in \cite{Haldane:1988zza}---namely time reversal and space inversion symmetry, while in the Haldane model only time reversal is broken whilst space inversion invariance is preserved. Different from \cite{Haldane:1988zza}, we do not propose band inversion but different gaps per valley, which is a crucial feature to allow for a chemical potential in the gap between the gaps.

\section{Half-integer topology behind the anomalous conductivity}
As we explained before, the anomalous conductivity \eqref{finaalresultaat} will receive no further quantum corrections, which is suggestive of a topological protection mechanism. That this is indeed the case will be scrutinized now. We lay the connection with a half-integer A-QHE in detail, not unlike the situation of the magnetically doped $(\textrm{BiSb})_2(\textrm{TeSe})_2$ materials reviewed in \cite{aqhe} albeit that the underlying mechanism is quite different in our case. Indeed, we are \emph{not} looking at the single Dirac cone per surface of a three-dimensional topological insulator, rather our judicious choice of model parameters, our genuine two-dimensional material eliminates one of the two cones from the game. Moreover, we do not have to rely on band inversion. What is common in both situations is that the effective low energy dynamics around the relevant Dirac point is attributable to a continuum Chern-Simons term, radiatively generated by the massive fermions.

For the benefit of the reader, let us briefly recapitulate the reasoning, see e.g.~\cite{Tong:2016kpv}. Reconsidering Eq.~\eqref{sigma}, and using the undisturbed energy eigenbasis, $\ket n$, and Hamiltonian time evolution with initial time $t=0$, $\vec j(t)= e^{i\frac{H_0t}{\hbar}} \vec j e^{-i\frac{H_0t}{\hbar}}$, we may rewrite it as
\be\label{sigma2}
\sigma = -\lim_{\omega\to0}\frac{1}{\hbar\omega} \int_{0}^{\infty} d\tau e^{i\omega\tau} \sum_n  \left(\bra{0} j_y\ket{n}\bra{n} j_x\ket{0}e^{\frac{i}{\hbar}(E_n-E_0)t}-\bra{0} j_x\ket{n}\bra{n} j_y\ket{0}e^{\frac{i}{\hbar}(E_0-E_n)t}\right).
\ee
Integrating over $\tau$ and expanding to leading order in $\omega$ again, we arrive at
\be\label{sigma2bis}
\sigma=2\hbar \text{Im}\sum_{n\neq0} \frac{\bra{0}j_y\ket{n}\bra{n}j_x\ket{0}}{(E_n-E_0)^2}.
\ee
No divergence in $\frac{1}{\omega}$ hampers the discussion as the photon self-energy (viz.~current correlator) starts at $\mathcal{O}(\omega)$, see \cite{Dudal:2018mms} for an explicit proof based on a Ward identity (in se gauge invariance).

Proceeding as in \cite{Tong:2016kpv} by introducing for both $\alpha$ filled bands and $\beta$ unfilled bands the Bloch wave functions, $\Psi^{\alpha,\beta}_{\vec p}(\vec{x}) = e^{i\vec p\cdot\vec x} u_{\vec p}^{\alpha,\beta}(\vec x)$, we can rewrite Eq.~\eqref{sigma2bis} using a compact notation
\be\label{sigma3}
\sigma=2\hbar \text{Im}\!\!\!\!\!\!\!\!\sum_{E_\alpha<\mu < E_\beta}\int_{T^2} \frac{d^2p}{(2\pi)^2} \frac{\bra{u_{\vec p}^\alpha}j_y\ket{u_{\vec p}^\beta}\bra{u_{\vec p}^\beta}j_x\ket{u_{\vec p}^\alpha}}{(E_\beta-E_\alpha)^2},
\ee
where the 2-dimensional $\vec p$ lives on the Brillouin zone $\sc T^2$. This expression can be further massaged into
\begin{eqnarray}\label{sigma4}
\sigma_{xy}&=&\frac{2e^2}{\hbar} \text{Im}\sum_{\alpha}\int_{T^2} \frac{d^2p}{(2\pi)^2} \braket{\partial_{p_y}u_{\vec p}^\alpha|\partial_{p_x}u_{\vec p}^\alpha}= \frac{e^2}{h} \sum_\alpha \mathcal{C}_\alpha
\end{eqnarray}
once the current in Fourier space is identified as $\vec j = \frac{e}{\hbar} \nabla_{\vec p} \widetilde H$, with $\widetilde H=e^{-i\vec p\cdot \vec x}He^{i\vec p\cdot \vec x}$, the Hamiltonian with momentum shifted over $\hbar \vec p$. To arrive at Eq.~\eqref{sigma4}, it was tacitly assumed that the bands are gap-separated, with chemical potential $\mu$ in such gap. The final outcome is the well-known sum over the integer Chern numbers $\mathcal{C}_\alpha$ per filled band \cite{kubo_Hall}, thereby disposing the topological character of the conductivity.

However, we should be careful when applying Eq.~\eqref{sigma4} directly to our case, as one of the underlying conditions is not met. Indeed, a crux of our setup is that the chemical potential $\mu$ was chosen in the gap between the 2 Dirac points. Evidently, this leads to a ``cutting'' of the respective band, see Fig.~1(d). Let us therefore analyze the expression \eqref{sigma3} a bit better in the case the underlying low energy effective Hamiltonian per valley per spin is of the type \cite{sticlet}
\be
H(\vec p) = h_0(\vec p)\mathbb{1}_2 + h_i(\vec p) \sigma_i.
\ee
The few materials that we will identify in the next section, with the desired gap structure, are all of this type. To be more precise, they are of the graphene-type, but supplemented with a specific mass term along the $\sigma^z$-direction.

Then, using the ``ray method'', it was shown in \cite{sticlet} that each massive Dirac point (cone) contributes with $\pm\frac{1}{2}$ ``topological charge''. If the full (smooth) band is taken into account for a double cone system (as we have), this will automatically lead to an integer Chern number and thence $\sigma = \pm \frac{e^2}{h}, 0$ as possible conductivities.

However, due to our judicious choice of $\mu$ in the finite intervalley gap between the $K$ and $K'$ Dirac points, we are effectively cutting away part of that band and only a single cone will contribute to the conductivity. Denoting with $\alpha'$ all completely filled bands and singling out that ``partially cut'' band (the encircled piece), we actually have
\begin{eqnarray}\label{sigma3}
\sigma_{xy}&=& \frac{e^2}{h} \underbrace{\sum_{\alpha'} \mathcal{C}_{\alpha'}}_{\mathrm{integer}} \pm\encircled{\frac{e^2}{h}\frac{1}{2}},
\end{eqnarray}
where the encircled piece is only there for the spin projection that is partially filled, the sign in front depending on which of the 2 Dirac points is the contributing one. Remember also that we lifted the spin degeneracy to avoid a net cancellation of the current; the other spin projection band is thus not filled and as such not contributing. This encircled piece exactly confirms the result \eqref{final} found before using an explicit quantum field theory computation and as promised, despite its half-integer character, the above reasoning shows it still carries a topological meaning. Indeed, small perturbations of the band geometry will preserve the fact that only a single Dirac point contributes to the Berry curvature integral, which always happens with a value of $\pm \frac{1}{2}$. This being said, the cutting of the band is a non-smooth operation, leading to a non-smooth integration zone, explaining why strictly speaking, we dot not find a topological integer. 

In future work, we will analyze the edge mode spectrum in full detail by solving the gapped Dirac equation associated to Eq.~\eqref{eq:chirallagrangian} for zero modes on a semi-infinite strip, but the judicious choice of the chemical potential in the gap between the different gaps, is expected to also lead to a halving of the number of (normalizable) edge modes, so the bulk-edge correspondence, \cite{nieuw4}, stills holds, thereby offering an alternative view on the origin of the half-integer conductivity and its robustness. Although in a different context~---and using different conventions, and not necessarily looking at zero modes---~\cite[Eq.~(5)]{nieuw3} already illustrates that the presence of a chemical potential can lead to one normalizable and one non-normalizable mode.

\section{Promising materials}

A few candidates present  the features that meet the criteria laid out in Section~II, as calculated using {\it ab initio} methods. In general, monolayer crystals with high potential to be applied in valleytronics physics are good candidates. First of all, manganese chalcogenophosphate $\rm{MnPSe}_3$ is organized in honeycomb lattices and it is intrinsically gapped. The material counts on two mechanisms to open the gaps: the antiferromagnetic order coupled to the sublattice and a spin-valley coupling, being responsible for time symmetry and inversion breaking, respectively \cite{antiferro1,antiferro2}. The situation corresponds exactly to the masses represented in Eq.~\eqref{eq:spin_coupling}. However, the spin remains degenerate due to  the antiferromagnetic (AFM) coupling between two inequivalent Mn ions.

\begin{figure}
	\includegraphics[width=0.8\textwidth]{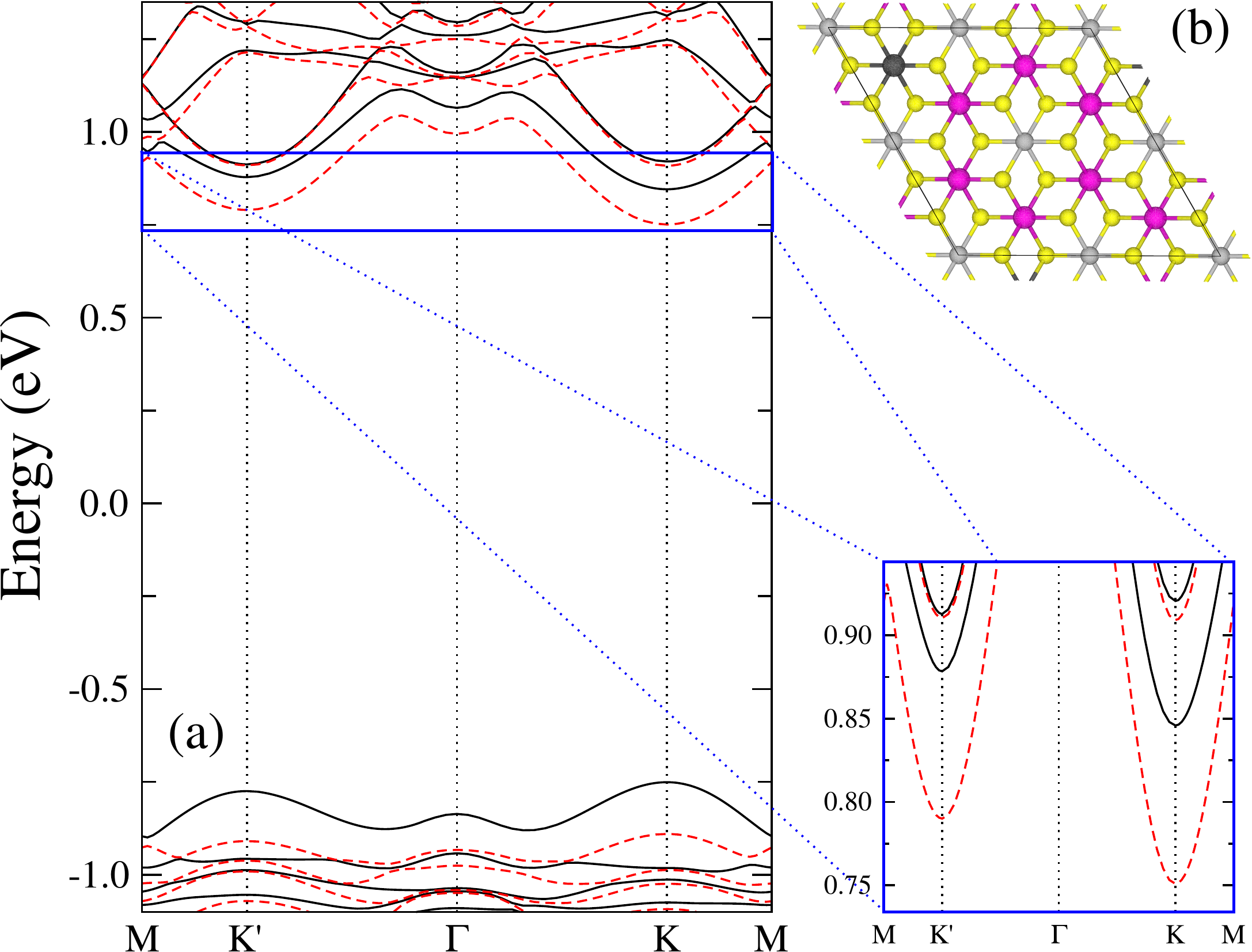}
	\caption{(a) DFT band structure of Zn-doped MnPSe$_3$. Spin up~(down) bands are represented by straight~(dashed) black~(red) lines. The inset shows the conduction band region in detail. (b) Unit cell of Zn-doped MnPSe$_3$ with pink, dark gray, yellow and gray spheres representing Mn, Zn, Se, and P, respectively. }
	\label{band_MnPSe3-Zn}
\end{figure}

In order to generate the necessary split between the spin bands, two types of setups can be prepared. One of them consists of doping the material, replacing for example one Mn atom by a Zn atom. The imbalance in spin will generate a magnetization, and consequently a spin splitting in the bands, while still retaining the different valleys \cite{zeeman}, as can be seen in Fig.~\ref{band_MnPSe3-Zn}. Note that we have the same structure schematically represented in Fig.~\ref{ABCDE}(e) with a spin up~(down) valley splitting of 33~(40)~meV and a spin splitting of 89~(93)~meV at the K'~(K) point. The band structure in Fig.~\ref{band_MnPSe3-Zn} was obtained via a density functional theory (DFT) calculation with the VASP package \cite{Kresse1996b} for the same system represented in \cite[Fig.~2(a)]{zeeman}. Exchange and correlation are described within the Perdew-Burke-Ernzerhof functional \cite{Perdew1996} with a on site Coulomb interaction of 5~eV applied to the Mn atoms (GGA~$+$~U) \cite{Dudarev1998}.

Another possible way is to use an heterojunction of different 2D materials such as MnPSe$_3$/CrBr$_3$ \cite{heterostructure}, MnPSe$_3$/MoS$_2$ \cite{heterostructure2} and  WS$_2$/$h$-VN \cite{heterostructure3}.

Besides the aforementioned heterostructures, dichalcogenides like NbSe$_2$ and WS$_2$ also seem promising to simultaneously lift both valley and spin degeneracy. This can be either reached by using a substrate \cite{dichalcogenide} to WS$_2$, or as by using a 2D magnetic semiconductor as monolayer  NbSe$_2$ with its large valley-polarized state \cite{2Dmagnetic2} . In such system, no external influence by either doping, substrating or van der Waals couplings is necessary, rather the magnetic state is sufficient to completely split the bands, resulting in the generation of a valley-polarized state with a spontaneously occurring valley current \cite{2Dmagnetic,2Dmagnetic1}. Our proposed effect thus is a specific type of a spin-polarized Quantum valley Hall effect, where the novel half-integer nature is due to a fine-tuning of the chemical potential. The latter can, for example, be realized in practice by applying a suitable external voltage.

\section{Conclusion}
We have proposed a half-integer version of the A-QHE, intrinsic to certain specifically gapped 2D materials, with as
{\it raison  d'\^{e}tre} the parity anomaly.

We offered several setups that follow the criteria to source this mechanism.  Therefore, we expect that our findings may open a new window to explore anomaly driven current in planar materials, offering potential advantages in the search for new anomalous transport phenomena.

Interestingly, for another class of transition metal chalcogenides, like V$_2$O$_3$, recent computations of \cite{V2O3,V2O3b} suggest that the two valleys contribute with \emph{equal} sign mass to the dynamical Chern-Simons term, something which should be related to the strong spin-orbit coupling. This could lead to an intrinsic integer rather than half-integer conductivity. This deserves further scrutiny in the future.

 \section*{Acknowledgments}
A.~J~.M.~receives partial support from FAPESP under fellowship number~2016/12705-7. F.~M.~thanks CNPq, Grant No.~381511/2018-9 and FAPESP, grant No.~2020/06896-0 for financial support. A.~R.~R.~was supported by FAPESP grants 2017/02317-2 and 2016/01343-7. C.~V. acknowledges financial support from FONDECYT under grants 1190192 and 1200483. We thank A.~Saraiva, I.~Shovkovy, S.~Mellaerts, L.~Levrouw and M.~Houssa for fruitful discussions.

\section*{Data availability}
Not applicable.

\section*{Competing interests}
The authors declare no competing interests.

\appendix

\bibliographystyle{apsrev4-2}

\end{document}